# Optimal Control of Time-Dependent Heat and Electromagnetic Energy Transfer in Selfdriven Coolers

Viorel Badescu

Candida Oancea Institute, Polytechnic University of Bucharest, Spl. Independentei 313,

Bucharest 060042, Romania;

Tel: 40.21.402.9339, Fax: 40.21.318.1019, Email: badescu@theta.termo.pub.ro

**Abstract**

The complex time-dependent heat and electromagnetic energy transfer in a new type of cooling system is analyzed. The system consists of a cold body, a Peltier element and an electric circuit containing an inductor with controllable inductance. This system allows cooling a body below the ambient temperature without using an external reservoir of mechanical work. The complex energy transfer in the new system is controllable. The minimum temperature reached by a cooled body depends on its mass. There is however a minimum minimorum temperature, for a specified mass of the cooled body. The minimum temperature of the cooled body decreases by increasing the thermal conductance of the Peltier element. Thermal damped oscillations may arise under special circumstances. This implies a very small difference between the initial temperatures of the cold and cooled body, respectively, and a specific range of variation for the conductance of the Peltier element.

1.INTRODUCTION

The Fourier law of heat conduction states that the heat flux between a warmer body and a colder body increases by increasing the difference of temperature between them. The thermal energy management is based on this law, with applications in many areas with societal impact such as space heating and the design of direct and reversed heat engines. The Fourier law continues to be a pillar of modern physics but recent technical advances appear to go beyond the popular way of presentation of this law. For instance, it has been shown that thermal diodes (rectifying the heat flux [1]) may be designed, based on phase change materials [2], quantum dots [3], metallic low temperature hybrid devices [4] and phononic devices [5].

These thermal diodes are the building blocks of more complex devices such as thermal transistors and logic gates [6].

It is well known that the second law of thermodynamics is formulated as (Clausius Statement): "Heat can never pass from a colder to a warmer body without some other change, connected therewith, occurring at the same time" [7]. In case (mechanical, electrical or chemical) work is entering the thermodynamic system, heat transfer from a lower temperature to a higher temperature is possible and this is the way that refrigerators and heat pumps are operating. According to the second law, the difference of temperature between a warmer body and a colder body placed in an isolated system will monotonically decrease until a thermal equilibrium is reached since heat does not flow by itself from a colder to a warmer body.

However, a complex heat transfer process has been recently imagined [8] allowing heat to flow from a colder body to a warmer body without mechanical work input but including a thermal inductor, first designed for heat capacity measurements [9]. It has been shown that the time evolution of this system does not break the second law and experimental results support this finding.

The purpose of this article is to analyze the complex heat and electromagnetic energy transfer in the new system and to show how this transfer may be controlled. The new system is called *selfdriven cooler* and opens perspectives for refrigeration without mechanical, electrical or chemical work input. Oscillatory temperature variations may occur during the operation of the new system, in special circumstances. Oscillations are often encountered in case of electrical quantities but they are rarely found in case of thermal quantities. These aspects are considered in this paper.

2. EXTERNALLY-DRIVEN AND SELFDRIVEN COOLERS

Assume a warmer body H and a colder body C. In case the two bodies are in direct thermal contact, heat is transferred from the body H towards the body C. The final temperature of the body H is higher or equal with the final temperature of body C, depending of the duration of the thermal contact, as the second law states. One calls *natural cooling* the process of decreasing the temperature of the warmer body H through heat transfer to the colder body C.

Assume one wants the final temperature of the body H to become lower than the final temperature of body C. By natural cooling the temperature of body H becomes equal with that of body C, after a long enough time interval. Farther decreasing the temperature of body H needs to continue extracting heat from this body but the second law forbids the natural transfer of that heat to the body C of higher temperature. However, heat may be extracted from the body H by a system consuming a specific form of work. The rate of the heat extracted depends on the rate of work consumed, which is an external factor since it may be chosen at will. The system consuming that work may be controlled from outside and is named *externally-driven cooler*.

The principle of a new type of cooling system is shown in Fig. 1. Assume a body H of temperature $T_H$ (mass $m_H$ and specific heat $c_H$) and a body C of temperature $T_C$ (mass $m_C$ and specific heat $c_C$) (Fig. 1). The assumption is that the initial temperature of the body H is higher than the initial temperature of the body C. The two bodies are in thermal contact with the two sides of a Peltier element of thermal conductance $k$ and combined Seebeck coefficient $\alpha$ of the used thermoelectric materials. Heat transfer is performed between the two bodies through the Peltier element. A closed electrical circuit includes the Peltier element (of internal resistance $R$) and an inductor of variable and controllable inductance $L$. The intensity of the electric current in this circuit is denoted $I$. When heat is transferred between the two bodies, a thermoelectric voltage is generated across the Peltier element and this induces an electric current in a closed circuit. Also, an electric current flowing through the Peltier element generates a positive or negative internal energy flux, depending on current direction.

The specificity is that the rate of heat extracted from the body H is selfdriven (in other words, it depends on the features of the heat transfer itself). However, in case the inductance $L$ is constant (see [8]) the heat rate extracted from the body H cannot be controlled from outside. The controllable inductance in Fig. 1 allows the heat rate extracted from the body H to be controlled from outside. The system consisting of the body C, the Peltier element and the electric circuit with a controllable thermal inductor is named *selfdriven cooler*.

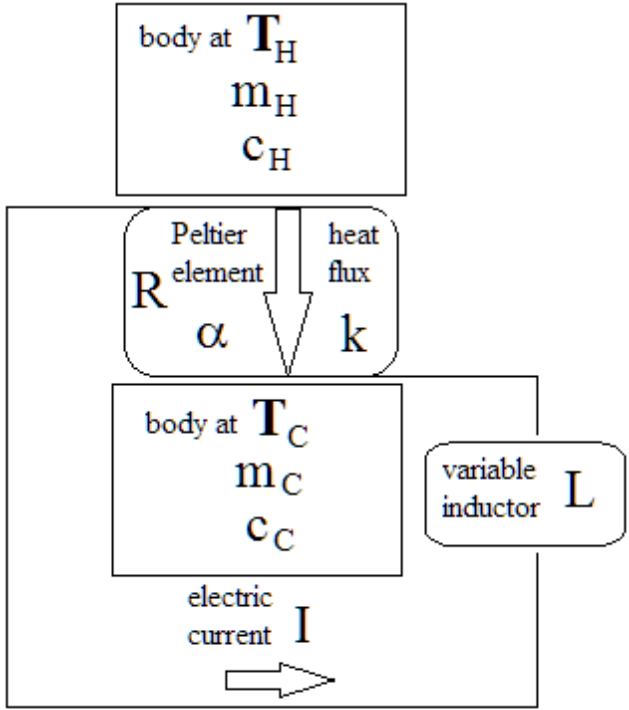

Figure 1. The system considered here. Heat is transferred between two bodies, of temperatures $T_H$ and $T_C$ (mass and specific heat $m_H$, $c_H$ and $m_C$, $c_C$, respectively). The heat flux passes through a Peltier element with thermal conductance $k$, combined Seebeck coefficient $\alpha$ of the used thermoelectric materials and internal electrical resistance $R$. The Peltier element is part of a closed electrical circuit, which includes a controllable inductor (of variable inductance $L$). The circuit is passes by an electric current of intensity $I$.

3. THEORY

3.1. Dynamics

The dynamics of the main quantities involved is shortly described next [8]. The time variation of the current intensity $I$ is governed by:

$$L\frac{dI}{dt} + RI - \alpha(T_H - T_C) = 0 \qquad (1)$$

In the l.h.s. of Eq. (1), the first term is the electromotive voltage coming from the Faraday's law of induction applied to the inductor of inductance $L$, the second term is the voltage drop due to the internal resistance $R$ and the third term is the thermoelectric voltage coming from the temperature difference between the two sides of the Peltier element.

The energy balance for the bodies of temperature $T_H$ and $T_C$ is given by, respectively:

$$m_H c_H \frac{dT_H}{dt} + \alpha T_H I - \frac{1}{2} R I^2 + k(T_H - T_C) = 0 \qquad (2)$$

$$m_C c_C \frac{dT_C}{dt} - \alpha T_C I - \frac{1}{2} R I^2 - k(T_H - T_C) = 0 \qquad (3)$$

In the l.h.s. of Eqs. (2) and (3), the first term is the time variation of the internal energy of the body, the second term is the internal energy flux absorbed/generated at the appropriate side of the Peltier element due to the flowing current, the third term is the heat generated by Joule effect in the Peltier element while the fourth term is the heat flux transferred by conduction from one body to the other body.

Several assumptions were used when writing Eqs. (2) and (3). The heat flux generated by Joule effect is assumed to be equally distributed between the two sides of the Peltier element while $\alpha$, $R$ and $k$ are taken as constants. The heat capacity of the Peltier element is neglected as well as other parasitic electrical or thermal effects. The heat flux is positive at input.

Equations (1-3) are solved by using the following initial conditions at time $t = 0$:

$$I(t=0) = 0 \qquad (4)$$

$$T_H(t=0) = T_{H,0} \qquad (5)$$

$$T_C(t=0) = T_{C,0} \qquad (6)$$

where $T_{H,0}$ and $T_{C,0}$ are known values. One assumes that $T_{H,0}$ is higher than $T_{C,0}$.

Equations (1) to (3) and the initial conditions Eqs. (4-6) develop the usual treatment of heat and electric charge transfer in Peltier elements [10] by including the electromotive voltage in Eq. (1). The theory is farther developed here by taking into account that the inductance is time dependent and controllable.

3.2. Optimal control problem

The optimization of the operation of selfdriven coolers requires defining the control and the objective function.

The control is the variable inductance $L$ which ranges between a minimum and a maximum value.

Three objective functions are considered in this paper. The minimization of the temperature $T_H$ at the end of the heat transfer process, at time $t_f$, is the first objective:

$$\text{Objective \#1: } T_H(t_f) \to min \qquad (7)$$

The second objective is the minimization of the duration $t_f$ of the heat transfer process necessary for the temperature of the initially warmer body to decrease from the initial value $T_{H0}$ to a given value $T_{Hgiven}$:

$$\text{Objective \#2: } t_f \xrightarrow[T_{H0} \to T_{Hgiven}]{} min \qquad (8)$$

The third objective is the maximization at the end of the heat transfer process (at time $t_f$) of the difference between the temperature of the initially colder body and the temperature of the initially warmer body:

$$\text{Objective \#3: } T_C(t_f) - T_H(t_f) \to max \qquad (9)$$

The optimal distributions of the temperatures $T_H(t)$ and $T_C(t)$ obtained by using the three objectives are different. Of interest in this paper is to decrease as much as possible the temperature of the initially warmer body. In this case the objective #1 is more relevant than the objective #3 since the extremized value of the difference $T_C(t_f) - T_H(t_f)$ depends on the variation of both $T_H(t)$ and $T_C(t)$. Therefore results for the objective #3 are not shown here.

In summary, the optimal control problem is defined as follows:

- *independent variable* : time $t$;

- *state variables*: current intensity $I$, temperature of body H, $T_B$, and temperature of body C, $T_C$;

- *objective* function: one of objective #1 or objective #2.

The objective function (either Eq. (7) or Eq. (8)) is minimized under the constraints of the ordinary differential Eqs. (1-3), which are solved by using the initial conditions Eqs. (4-6).

3.3. Direct optimal control procedure

Equations (1-3) and, as a consequence, the Hamiltonian of the optimal control problem, are linear in the inverse of the control function $(1/\hat{L})$. Therefore, a non regular solution is to be expected. The optimal control is either singular or of the bang-bang (all or nothing) type, depending on problem constraints [11]. Indirect optimal control methods (such as the Maximum Principle of Pontryagin) may be used to obtain the switching structure. This requires building a system consisting of Eqs. (1-3) and their adjoint equations, as well as defining appropriate boundary conditions [11]. Direct optimal control methods do not need building adjoint equations and they are less dependent on the details of the initialization stage. Therefore, they have wide usage in engineering applications. A direct optimal control method is used here.

The principle of the direct optimal control method is shortly described next. The dynamics of the state variables and controls is described by several ordinary differential equations and constitutes the constraint under which the objective function is minimized or maximized in the optimal control problem (OCP). The OCP is transformed into a non-linear problem (NLP). First, the independent variable is discretized. Next, this discretization is applied to the state and control variables as well as to all constraints, including the ordinary differential equations. For details on NLP optimization and direct transcription methods see e.g. [12,13].

The open-source programming package BOCOP [14] is used here. BOCOP implements a direct optimal control method. The user defines by using C++ programming language the ordinary differential equations describing the dynamics of the problem, the objective function and additional constraints on state and control variables. The transcription of the OCP into a NLP is performed internally by BOCOP algorithms. The NLP is solved by using the IPOPT solver [15] which is based on a primal-dual interior point algorithm. During the optimization process the automatic differentiation tool ADOL-C [16] is used to perform the derivatives.

BOCOP is built for the *minimization* of the objective function. In case the objective is to maximize a function (such as in case of Eq. (9)), BOCOP requires defining a new objective (i.e. $-[T_C(t_f) - T_H(t_f)] \to min$). BOCOP has several discretization method options. We used in most cases the option Midpoint (implicit, 1- stage, order 1). The number of discretization steps for the independent variable was 1000. The maximum allowed number of iterations was 10,000 while the tolerance is $10^{-14}$.

## 4. UNCONTROLLED COOLING

We shortly investigate the uncontrolled cooling process [8]. Therefore, the inductance $L$ of the inductor is kept constant in time. The results may be used as a reference when the operation of a selfdriven cooler is analyzed.

A few comments follow about the magnitude of the inductance encountered in practice. Inductors with inductances up to 20 H are presently commercialized. Measured inductances for heating inductors [17] and melting inductors [18] have similar order of magnitude. Inductors with much higher inductance values have been built for special destinations. For instance, inductances of the order of 1000 H are quoted in relation with the Tevatron magnet of the Fermi National Accelerator Laboratory and high power transformers [19]. The inductance values used in [8] range between 30 H and 150 H. Most cases presented in this paper involve inductance values in this range. Inductances outside this range are however assumed in few cases, to give perspective to the results.

Table 1. Values adopted for several quantities unless other values are specified.

| **Reference values** | |
|---|---|
| Time, $t_{ref}$ (s) | 300 |
| Current intensity, $I_{ref}$ (A) | 0.4 |
| Temperature, $T_{ref}$ (K) | 373.15 |
| Inductance, $L_{ref}$ (H) | 150 |
| **Parameters** | |
| Seebeck coefficient, $\alpha$ (V/K) | $3.21 \cdot 10^{-3}$ |
| Thermal conductance, $k$ (W/K) | 0.0318 |
| Electric resistance, $R$ ($\Omega$) | 0.22 |
| Initially warmer body mass, $m_H$ (kg) | 0.009 |
| Initially colder body mass, $m_C$ (kg) | 3.65 |
| Specific heat of the two bodies (copper), $c_H = c_C$ (J/(kgK)) | 390 |

Table 1 shows the values adopted in this paper for the main quantities. The parameter values are identical with those used in [8] for uncontrolled cooling processes. These values are unchanged (except other values are explicitly stated). Constant inductance values between 60 H and 150 H were used in this work. In addition, a high conductance value $L = 750\ H$ has been used, to give perspective to the results.

The time evolution of the system has been analyzed until the end of the heat transfer process at time $t_f = 1200\ s$.

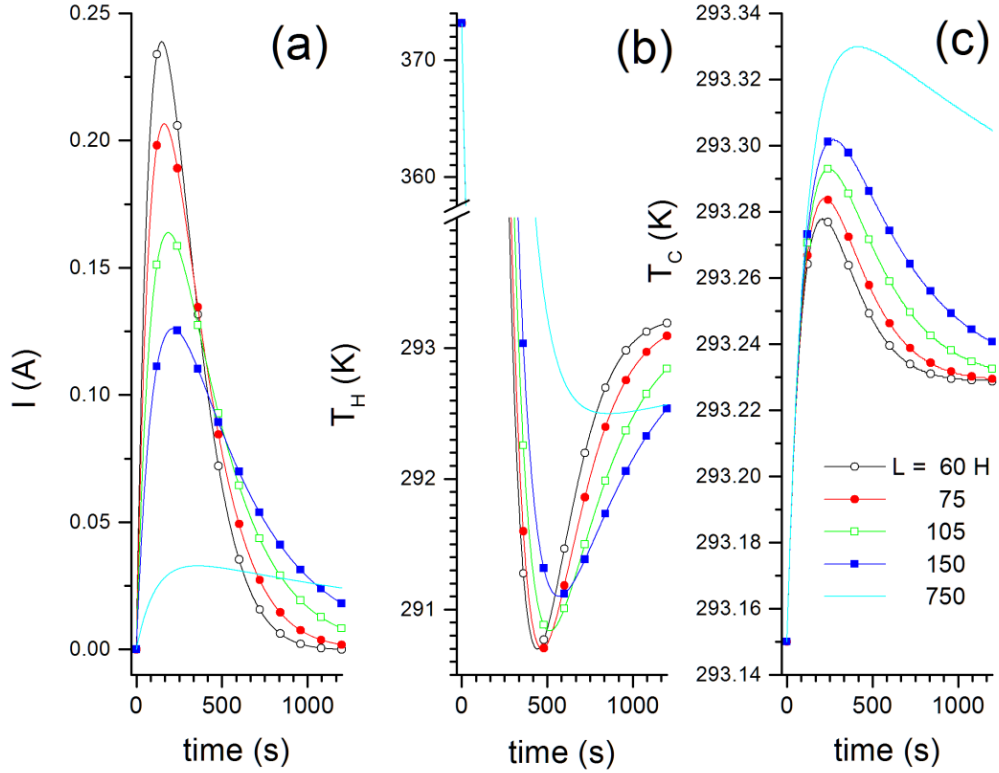

Figure 2. Time dependence of (a) current intensity $I$, (b) temperature $T_H$ of the initially warmer body and (c) temperature $T_C$ of the initially colder body. Different values of the constant inductance $L$ have been considered.

The current intensity $I$ and the temperatures $T_H$ and $T_C$ have a similar time variation, for all values of the inductance (Figure 2a,b,c). $I$ has a maximum, which is higher for smaller values of $L$, followed by a tail, which is longer for larger values of $L$ (Fig. 2a). The time when the maximum value of $I$ occurs slightly depends on the inductance value. The temperature $T_H$ is of major interest in this work. Starting from the initial time $t = 0\ s$, $T_H$ decreases and reaches a minimum value $T_{H,min}$ at time $t_{T_{H\,min}}$ (Fig. 2b). Both $T_{H,min}$ and $t_{T_{H\,min}}$ depend on the inductance value. Next, the $T_H$ increases and reaches the value $T_H(t_f)$ at the end of the heat transfer process at time $t_f$. The temperature $T_C$ increases from the initial value $T_{C0}$ at time $t = 0\ s$, reaches a maximum and next decreases (Fig. 2c). The range of variation of $T_C$ is very small, less than 0.2 °C, for all inductance values.

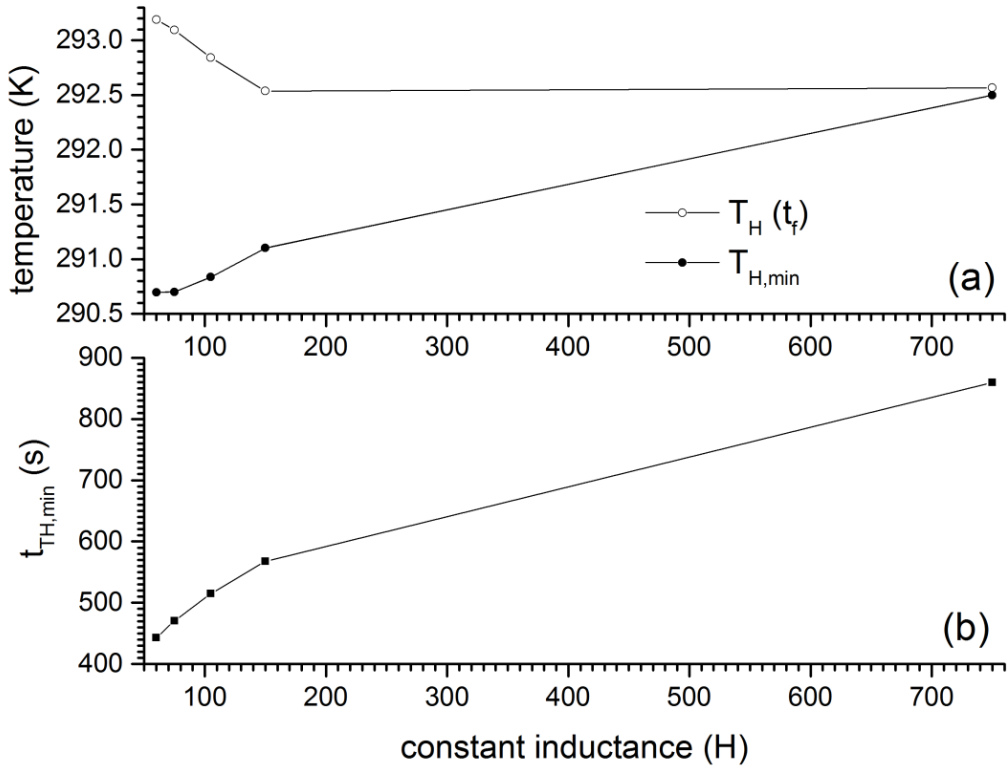

Figure 3 Dependence of (a) temperatures $T_{H,min}$ and $T_H(t_f)$ and (b) time $t_{T_{H\ min}}$ of the minimum temperature $T_{H,min}$, on the value of the constant inductance $L$.

Figure 3 shows the dependence of $T_{H,min}$, $T_H(t_f)$ and $t_{T_{H\ min}}$ on the value of $L$. The minimum temperature $T_{H,min}$ of the initially warmer body is lower that 292.5 K, for all inductance values considered here (Fig. 3a). This means that the initially warmer body is cooled down to a temperature below that of the initially colder body, which is higher than 293.15, for all inductance values (See Fig. 2c). This agrees with results reported in [8].

$T_{H,min}$ is quite constant for smaller inductance values and increases by increasing $L$, at higher inductance values (Fig. 3a). Therefore, an important observation is that enhancing the heat transfer effect does not necessarily need high inductance values. The temperature $T_H(t_f)$ at the end of the heat transfer process decreases by increasing the inductance value, at low values of $L$, but it is weakly dependent on $L$ at high inductances. Generally, the difference $T_H(t_f) - T_{H,min}$ decreases at larger inductance values. The time $t_{T_{H\,min}}$ when $T_H$ is a minimum is a critical quantity since the heat transfer process should be stopped at this time to obtain maximum cooling effects. $t_{T_{H\,min}}$ increases by increasing the value of $L$ (Fig. 2b). Therefore, faster cooling effects are obtained at smaller constant inductance values.

5. OPTIMAL OPERATION OF SELFDRIVEN COOLERS

In this section we explore the potential of the new cooling technique based on selfdriven coolers. The main parameters have the values shown in Table 1 and they are kept constant unless otherwise specified. The optimal operation of the selfdriven cooler is affected by several factors, as described next.

5.1. Duration of the cooling process

Results obtained for the optimal operation of the selfdriven cooler are presented in Fig. 4. The range of variation of the control (i.e. the variable inductance $L$) is from 30 H to 150 H while the duration of the heat transfer process is $t_f = 1000\ s$. The optimal control is of bang-bang type and has a jump from the lowest to the highest allowed values at time $t_{jump} = 107\ s$ (Fig. 4d). The time variation of the current intensity $I$ and temperatures $T_H$ and $T_C$ has a similar shape with the time variation of these quantities in case of uncontrolled heat transfer (compare Fig. 4a,b,c with Fig. 2a,b,c, respectively). Figure 4b has a break to show that $T_H$ has a minimum value of 287.20 K at time $t_{T_{H\,min}} = 471\ s$. This minimum temperature $T_{H,min}$ is lower for controlled heat transfer than for uncontrolled heat transfer, as expected (compare Fig. 4b with Fig. 2b and Fig. 3a). There is a drop of about three degrees centigrade when the heat transfer is controlled.

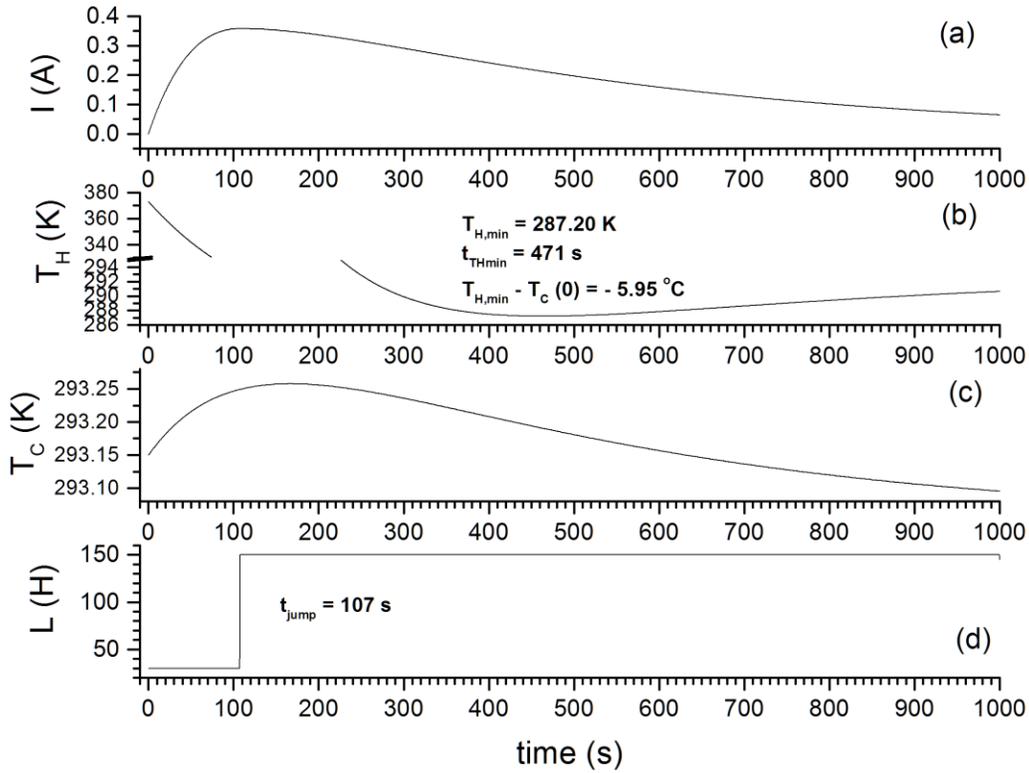

Figure 4. Optimal time variation of several quantities in case the control $L$ ranges between 30 H and 150 H. (a) intensity of electric current $I$ ; (b) temperature $T_H$ of the initially warmer body; (c) temperature $T_C$ of the initially colder body; (c) inductance $L$ .

Heat transfer processes of duration $t_f$ up to 3000 s have been considered here (see Fig. 5). The range of variation of the control (i.e. the variable inductance $L$) is from 30 H to 150 H . For processes shorter than $t_f = 429\ s$ there is no control jump (see Fig. 5a). The control is constant during the process at the minimum value $L_{min} = 30H$ . All heat transfer processes longer than $t_f = 429\ s$ have a control jump from $L_{min} = 30H$ to the maximum value $L_{max} = 150H$ at the same time $t_{jump} = 107\ s$ . The time when the minimum value $T_{H,min}$ is reached, $t_{T_{H\ min}}$ , increases by increasing the process duration $t_f$ but for processes longer than (roughly) 500 s it is the same (Fig 5a). The control jump is always anterior to the moment when $T_{H,min}$ is reached, except for processes shorter than 429 s. The main observation is that for sufficiently longer heat transfer processes $t_{jump}$ and $t_{T_{H\,min}}$ do not depend on process duration $t_f$. For rather short processes the minimum temperature $T_{H,min}$ is reached at the end of the process (Fig. 5b). This means that $T_{H,min} = T_H(t_f)$. Longer heat transfer processes are characterized by $T_{H,min} < T_H(t_f)$ and $T_{H,min}$ may be smaller that $T_H(t_f)$ by more than 4 degrees centigrade. However, for very long processes the difference between the two quantities becomes a constant.

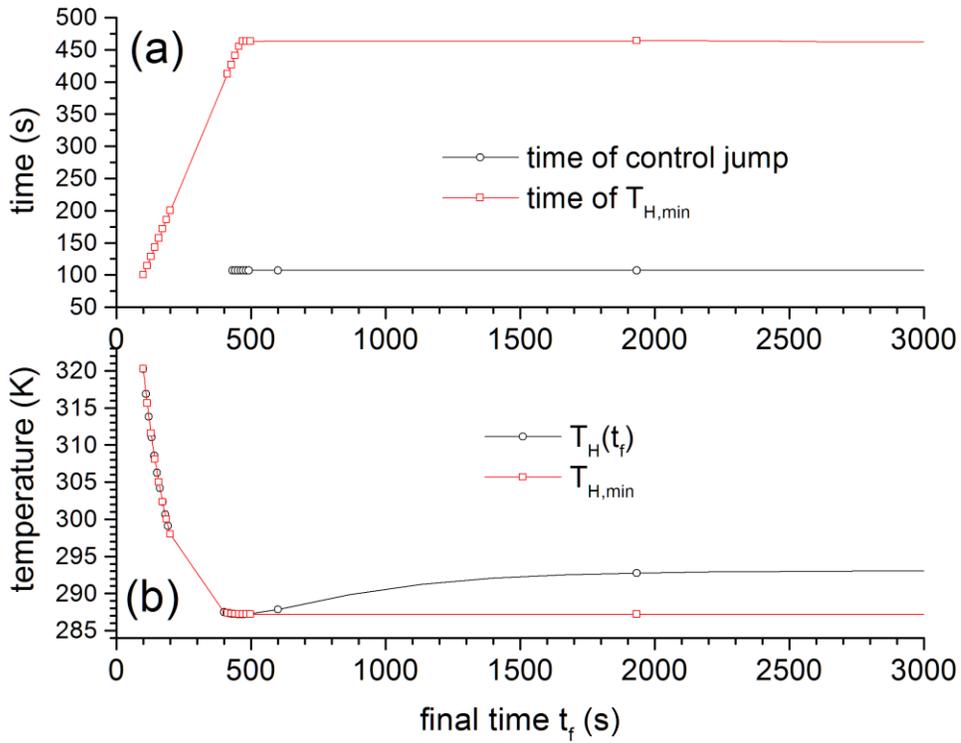

Figure 5. Dependence of several quantities on the duration of the heat transfer process, $t_f$. (a) time of the control jump, $t_{jump}$ and time of the minimum $T_H$ temperature; (b) minimum $T_H$ temperature and temperature $T_H$ at the end of the heat transfer process.

5.2 Domain of the variable inductance

One expects the range of variation of the control values $[L_{min}, L_{max}]$ plays an important role in the controlled heat transfer process. Table 2 shows results for controlled heat transfer processes. First row of Table 2 shows results for an uncontrolled heat transfer process, to be used as a reference. To give perspective, results corresponding to very large inductance values, outside the common range, are shown in the last three rows of Table 2.

For longer enough duration $t_f$ of the heat transfer process the switching scheme consists of a control jump between $L_{min}$ and $L_{max}$ (see Fig. 5a). The jump time $t_{jump}$ depends on the range of variation of the control values $[L_{min}, L_{max}]$. Generally, $t_{jump}$ increases by increasing $L_{min}$ and $L_{max}$. At very large values of $L_{max}$ the jump time becomes a constant (see last two rows in Table 2).

The temperature $T_H$ at the final time $t_f$ of the heat transfer process increases by increasing $L_{min}$ (at constant $L_{max}$). However, $T_H(t_f)$ does not depend in a systematic way on $L_{max}$ (at constant $L_{min}$). For instance, $T_H(t_f)$ decreases by increasing $L_{max}$ (at small $L_{max}$ values) and increases by increasing $L_{max}$ (at large $L_{max}$ values).

The optimal control strategy provides better results than uncontrolled heat transfer processes, as expected. Indeed, the temperature $T_{H\,min}$ for the uncontrolled heat transfer process in row 1 of Table 2 is larger than that of all controlled processes in that table. Generally, $T_{H\,min}$ decreases by increasing the width of the interval $[L_{min}, L_{max}]$.

The time when temperature $T_{H\,min}$ is reached increases by increasing $L_{max}$ (at constant $L_{min}$). The dependence of $T_{H\,min}$ on $L_{min}$ (at constant $L_{max}$) is not monotonous (see rows 3 and 4 in Table 2).

Table 2. Dependence of several quantities on the range of variation of the control values $[L_{min}, L_{max}]$. $t_{jump}$ - jump time; $T_H(t_f)$ - temperature $T_H$ at the final time $t_f$ of the heat transfer process; $T_{H\,min}$ - minimum temperature $T_H$; $t_{T_{H\,min}}$ - time when temperature $T_{H\,min}$ is reached. $t_f = 1000\ s$.

| $L_{min}$ (H) | $L_{max}$ (H) | $t_{jump}$ (s) | $T_H(t_f)$ (K) | $T_{H\,min}$ (K) | $t_{T_{H\,min}}$ (s) |
|---|---|---|---|---|---|
| 60 | 60 | none | 292.84 | **290.70** | 470 |
| 30 | 150 | 107 | 290.64 | 287.20 | 464 |
| 75 | 150 | 162 | 291.59 | 289.72 | 518 |
| 105 | 150 | 186 | 291.88 | 290.45 | 514 |
| 30 | 105 | 232 | 288.05 | 288.01 | 945 |
| 30 | 75 | 107 | 292.65 | 288.85 | 416 |
| 75 | 135 | 162 | 291.82 | 289.85 | 510 |
| 60 | 120 | 148 | 291.92 | 289.51 | 487 |
| 15 | 1500 | 75 | 281.27 | 280.88 | 663 |
| 1.5 | 1500 | 18 | 272.64 | 271.93 | 610 |
| 1.5 | 7500 | 18 | 270.50 | not found | >1000 s |

5.3 Mass of the cooled body

Optimally controlled heat transfer processes with inductance ranging between $L_{min} = 30\ H$ and $L_{max} = 150\ H$ have been considered for different masses $m_H$ of the body H. The duration of the process changes with the $m_H$ value, as follows: for $m_H \leq 0.01\ kg$, $t_f = 1000\ s$; for $0.02\ kg \leq m_H \leq 0.04\ kg$, $t_f = 2000\ s$; for $m_H = 0.05\ kg$, $t_f = 2500\ s$ while for $m_H = 0.1\ kg$, $t_f = 6000\ s$. This variable $t_f$ value is needed to allow finding the time $t_{T_{H\,min}}$ when the minimum temperature $T_{H\,min}$ is reached.

The switching scheme is similar in all cases, with a control jump from $L_{min}$ to $L_{max}$ (see Fig. 4d, for instance). The time of the control jump, $t_{jump}$, is longer for larger values of $m_H$ (Fig. 6a). The optimal time variation of temperature $T_H$ has the usual shape of Fig. 4b. The time when the minimum value of the temperature $T_H$ is reached, increases by increasing the mass $m_H$ (Fig. 6a). The control jump is always anterior to the occurrence of the minimum temperature $T_{H\,min}$, for the same mass $m_H$ (i.e. $t_{jump} < t_{TH\,min}$).

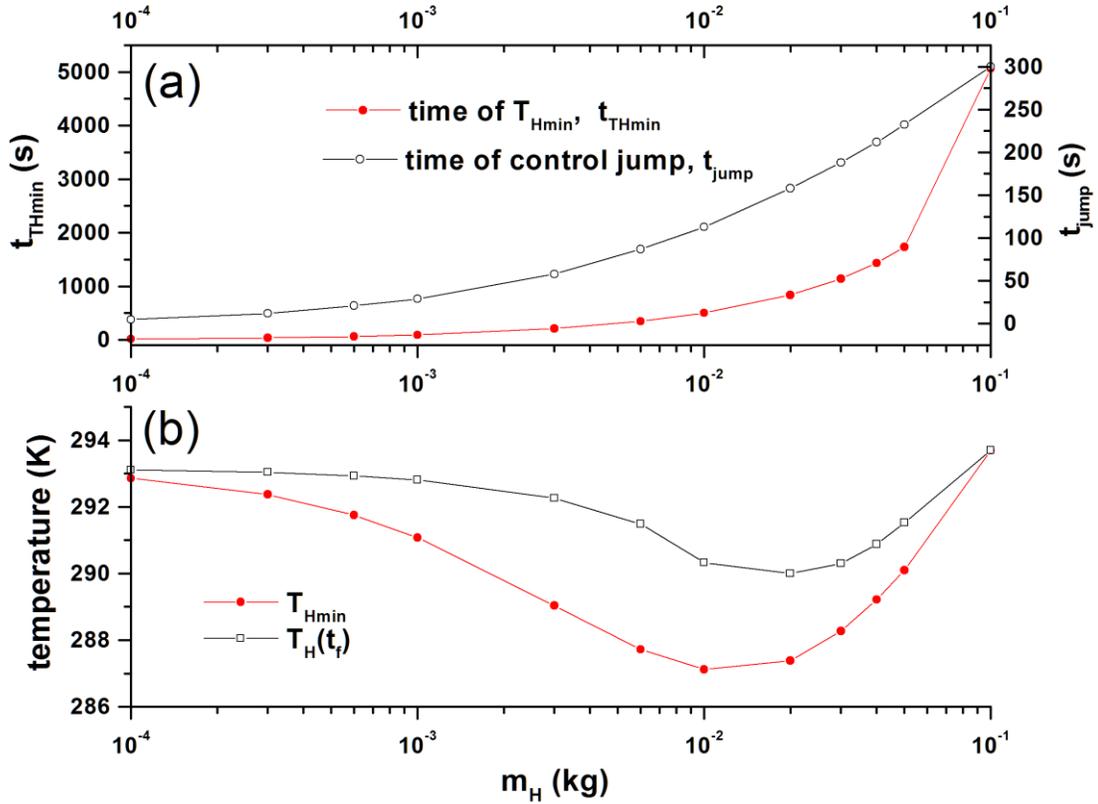

Figure 6. Dependence of several quantities on the mass of the cooled body H. An optimal heat transfer process with inductance ranging between $L_{min} = 30\,H$ and $L_{max} = 150\,H$ has been considered. (a) time of control jump and time of minimum $T_H$ temperature; (b) minimum temperature $T_{H\,min}$ and temperature of the H body at the final time of the process, $T_H(t_f)$

The minimum temperature $T_{H\,min}$ depends on the mass of the body H, as shown in Fig. 6b. $T_{H\,min}$ decreases by increasing the mass $m_H$ and reaches a *minimum minimorum* value for an optimum mass $m_{Hopt}$ between 0.01 kg and 0.02 kg. This is a remarkable result which puts constraints on the mass of the body H to be effectively cooled in practice. For $m > m_{Hopt}$, $T_{H\,min}$ increases by increasing the mass. The time needed to reach the temperature $T_{H\,min}$ is longer that one hour in case of $m_H = 0.1\,kg$ (Fig. 6a). The temperature of the body H at the end of the process, $T_H(t_f)$, depends on $m_H$ (Fig. 6b). $T_H(t_f)$ has a minimum for $m_H$ between 0.01 kg and 0.02 kg.

Assume that a proper optimum value of the mass $m_H$ has been identified by using objective function #1, as shown in Fig. 6b. The objective function #1 minimizes the value $T_H(t_f)$ of the body H temperature at the end of the heat transfer interval and an associated minimum temperature $T_{H\,min}$ is identified. However, the objective function #1 is not to find that specific temperature value $T_{H\,min}$. The minimum heat transfer time to obtain the precise value $T_{H\,min}$ may be obtained by using the objective function #2. This is shortly investigated next.

The objective function #2 defined by Eq. (8) estimates the minimum time necessary to cool down the temperature of the body H from a given initial temperature $T_{H0}$ to a given temperature $T_{Hgiven}$. A solution for this optimal control problem does not always exist. There is no control strategy to cool down the body H to a given temperature $T_{Hgiven}$ if this one is too low as compared to the initial temperature $T_{H0}$. However, if such a control strategy exists, then its switching scheme is similar with that shown previously. It consists of a control jump from the minimum value $L_{min}$ to the maximum value $L_{max}$ (see Fig. 4d, for instance). The time variation of the temperature $T_C$ has a maximum similar with that shown in Fig. 4c. However, in case of the objective function #2 (Eq. (8)) the optimal variation of the temperature $T_H$ is different from that obtained in case of the objective function #1 (Eq. (7)), which always exhibits a minimum value $T_{H\,min}$ (see Fig. 4b, for instance). In case of the objective function #2 the temperature $T_H$ monotonously decreases from the value $T_{H0}$ to the value $T_{Hgiven}$.

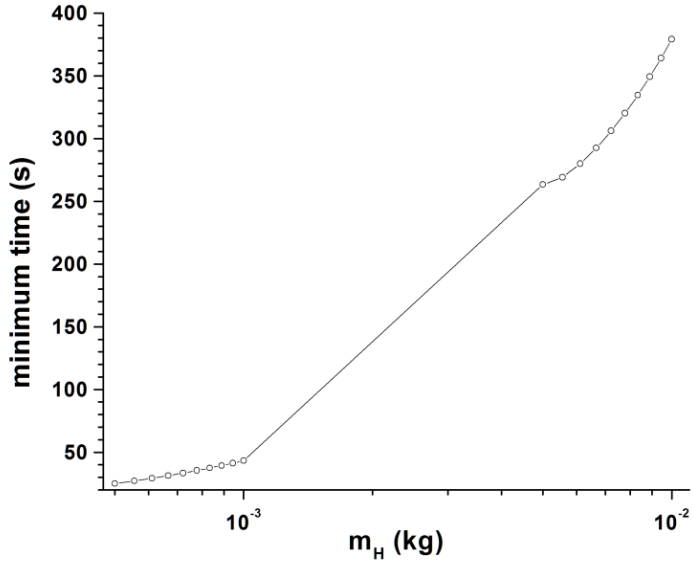

Figure 7. The minimum time $t_{f,min}$ necessary to cool down the temperature of the body H from $T_{H0} = 373.15\ K$ to $T_{H,given} = 288.15\ K$ as a function of the mass $m_H$ of the body H. Initial temperature of body C is $T_{C0} = 293.15\ K$. The control is ranging between $L_{min} = 30\ H$ and $L_{max} = 150\ H$.

The minimum time to cool down the temperature of the body H to a given temperature $T_{H,given}$ increases by increasing the mass of the body H, as expected (Fig. 7).

5.4 Properties of the Peltier element

It is expected that the material properties, i.e. the Seebeck coefficient $\alpha$ and the conductance $k$, have significant influence on the heat transfer process. The authors of Ref. [8] used Peltier elements produced by Kryotherm Inc. (i.e. module TB-7-1.4-2.5, standard single stage; thermoelectric coolers). Information about the properties $\alpha$ and $k$ for other Peltier elements produced by Kryotherm Inc. are not explicitly available for public usage. Therefore we access information for Peltier elements produced by TEC Microsystems. The information available for TEC Mycrosystems coolers does not allow a simple estimation of their properties $\alpha$ and $k$. However, the information available for TEC Mycrosystems generators allows easier computation of the properties $\alpha$ and $k$ [20]. These properties are estimated simultaneously since a Peltier element is characterized by a specific couple of values for $\alpha$ and $k$. Table 3 shows the values of $\alpha$ and $k$ for four elements.

Table 3. Values of Seebeck coefficient $\alpha$ and thermal conductance $k$ for four TEC Mycrosystems elements.

| Element Number | Element type | $k$ (W/K) | $\alpha$ (V/K) |
|---|---|---|---|
| 1 | 1MC04-007-15-TEG | $2.22 \cdot 10^{-3}$ | $2.8 \cdot 10^{-3}$ |
| 2 | 1MC04-070-015-TEG | $22.2 \cdot 10^{-3}$ | $28 \cdot 10^{-3}$ |
| 3 | 1MC06-030-05-TEG | $62.8 \cdot 10^{-3}$ | $12.0 \cdot 10^{-3}$ |
| 4 | 1MC06-126-05-TEG | $263 \cdot 10^{-3}$ | $50.4 \cdot 10^{-3}$ |

The four elements #1 to #4 in Table 3 are ordered according to their increasing thermal conductance. The optimal control for the element with the lowest conductance (#1) switches two times, from the value $L_{max}$ to $L_{min}$ and vice-versa (Fig. 8b). The optimal control for the other three elements (#2 to #4) have just one jump, from $L_{min}$ to $L_{max}$ (Figs. 8d,f,h, respectively). The switching time $t_{jump}$ decreases by increasing the value of the thermal conductance (compare Figs. 8d,f,h). The time variation of the temperature $T_H$ for the element with the lowest conductance (i.e. #1) is smooth, with a very shallow minimum (Fig. 8a). The time variation is more abrupt in case of element #2 (Fig. 8c) and much more abrupt for the elements #3 and #4, with the highest thermal conductance (Figs. 8e and 8g, respectively). The minimum temperature $T_{H\,min}$ decreases by increasing the thermal conductance (compare Figs. 8a,c,e,g). In case of the elements #3 and #4, $T_{H\,min}$ is significantly smaller than the initial temperature of the body C (Figs. 8e,g). The time when the minimum temperature $T_{H\,min}$ is reached decreases by increasing the thermal conductance (compare Figs 8a,c,e,g) and it is always posterior to the time of the control switching (i.e. $T_{H\,min} > t_{jump}$, for given element).

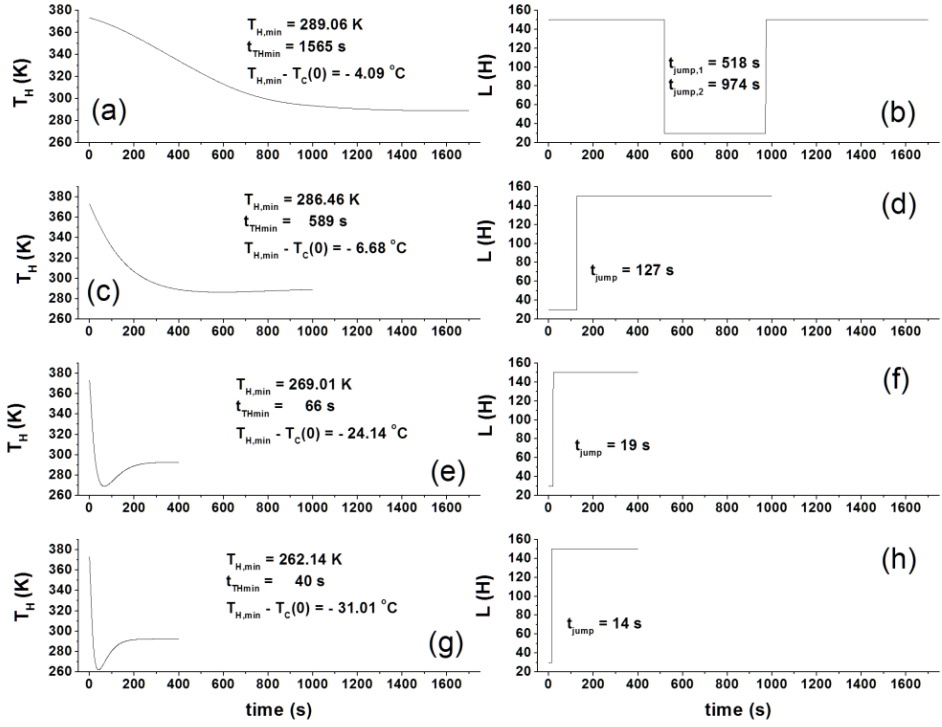

Figure 8. Time variation of the temperature $T_H$ (a,c,e,g) and control $L$ (b,d,f,h) for optimally controlled heat transfer process. The four Peltier elements of Table 3 were considered. (a,b) – element #1; (c,d) – element #2; (e,f) – element #3; (g,h) – element #4. $L$ - inductance ranging between $L_{min} = 30\ H$ and $L_{max} = 150\ H$.

5.5. Thermal oscillations

The analysis performed in [8] is based on several approximations and simplification of the dynamic differential equations. A remarkable result is that an oscillatory uncontrolled heat transfer process may be obtained in special circumstances. For instance, such regime is obtained in for $R = 0.22\ \Omega$, $k = 0.0318\ W/K$, $m_C c_C = 1423.5\ J/K$ and the very high inductance $L^* \equiv RC_C / k = 9848.113\ H$. The conclusion is that a necessary condition for the occurrence of an oscillatory regime is:

$$\Delta_0 \equiv T_{H0} - T_{C0} << T_{C0} \qquad (10)$$

This condition corresponds to an initially colder body of very large mass whose temperature $T_{C0}$ is constant in time. Therefore this phenomenon deserves farther investigations, under a proper, not-simplified, treatment.

Here both temperatures $T_H$ and $T_C$ are time dependent and the condition Eq. (10) turns into:

$$\Delta_0 \equiv T_{H0} - T_{C0} << \frac{T_{H0} + T_{C0}}{2} \qquad (11)$$

Simplifications of the differential equations are not adopted here. A guess and check procedure has been used to find parameter values associated with an uncontrolled oscillatory regime. The condition Eq. (11) has been also used.

The following parameter values have been used: $\alpha = 0.00321 \ V/K$, $k = 0.2 \cdot 10^{-6} \ W/K$, $L = 150 \ H$, $T_{C0} = 293.15 \ K$. For $\Delta_0 = 2 \ K$ the condition Eq. (11) is fulfilled but an oscillatory heat transfer is not obtained. However, $\Delta_0 = 0.2 \ K$ obeys Eq. (11) and is associated with the uncontrolled damped oscillatory heat transfer process shown in Fig. 9. The electrical current intensity changes during the time from positive to negative values and vice-versa (Fig. 9a). This is associated with an alternation of heat generation and heat absorption in the Peltier element, respectively. The temperature $T_H$ is initially higher than the temperature $T_C$ (see Fig. 9b and Fig. 9c) and this corresponds to heat transfer from H body to the C body. However, at time $t_{T_{H \ min}} = 1376 \ s$, $T_H$ is lower that $T_C$ and heat is transferred from the C body to the H body. Later on the temperature $T_H$ increases and the heat transfer changes again its direction.

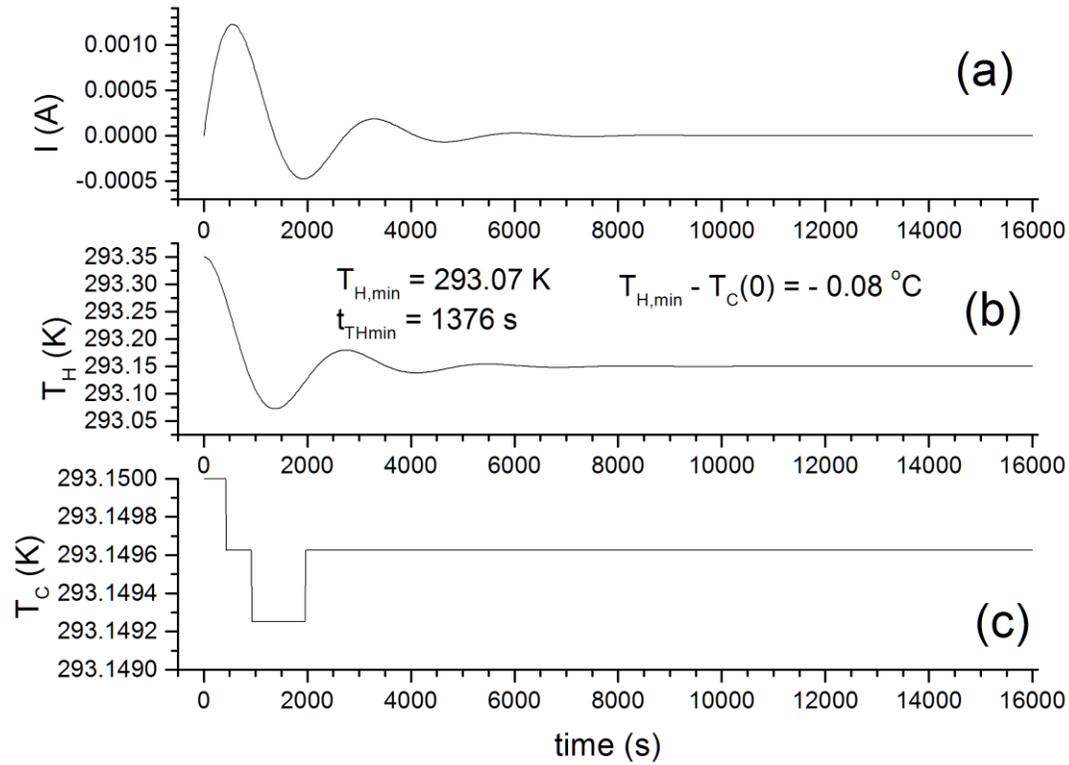

Figure 9. Oscillatory time variation of several quantities in case of uncontrolled heat transfer. (a) Electrical current intensity $I$; (b) temperature $T_H$ of the initially warmer body; (c) temperature $T_C$ of the initially colder body. Constant inductance $L = 150\ H$, Seebeck coefficient $\alpha = 0.00321\ V/K$, conductance $k = 0.2 \cdot 10^{-6}\ W/K$, initial temperature $T_{C0} = 293.15\ K$, $\Delta_0 \equiv T_{H0} - T_{C0} = 0.2\ K$. Duration of the heat transfer process $t_f = 16000\ s$.

The question we face now is if a controlled oscillatory heat transfer process is possible. We assume a controllable inductance in the range $L_{min} = 75\ H$ to $L_{max} = 225\ H$ and the duration of the process is $t_f = 16000\ s$. The parameter values are the same as in the case of the uncontrolled heat transfer process described in Fig. 9 but several values of the conductance $k$ are considered. Results are shown in Fig. 10.

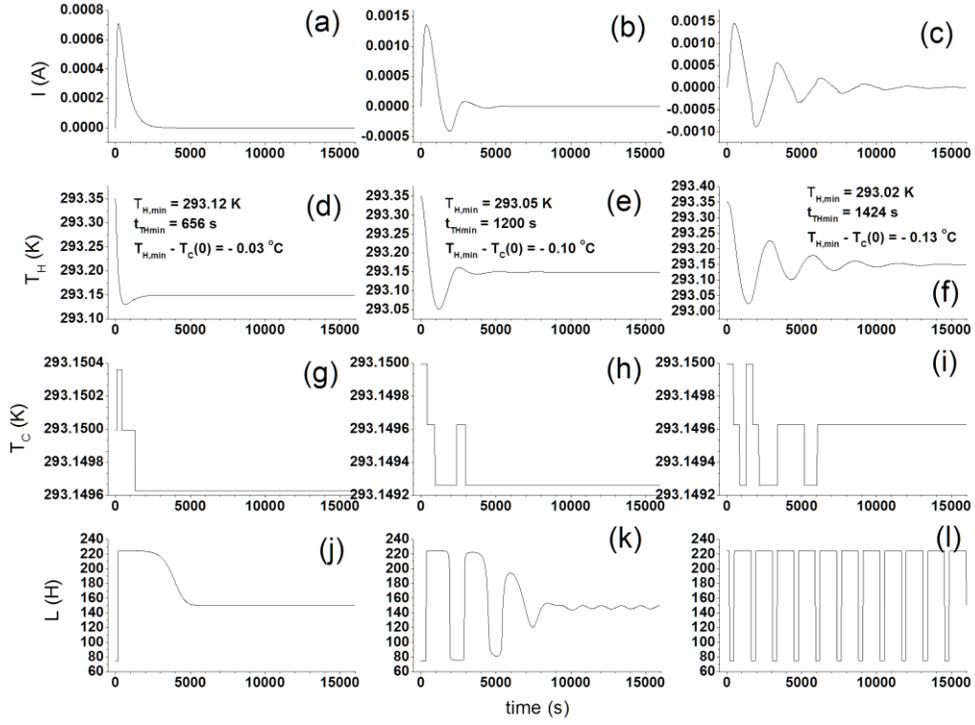

Figure 10. Time variation of several quantities for three optimally controlled heat transfer processes. (a,d,g,j) conductance $k = 2 \cdot 10^{-2}\ W/K$; (b,e,h,k) $k = 2 \cdot 10^{-3}\ W/K$; (k,f,i,l) $k = 2 \cdot 10^{-4}\ W/K$. $I$ - electric current intensity; $T_H$ - temperature of the body H; $T_C$ - temperature of the body C; $L$ - inductance ranging between $L_{min} = 75\ H$ and $L_{max} = 225\ H$. Seebeck coefficient $\alpha = 0.00321\ V/K$, initial temperature $T_{C0} = 293.15\ K$, $\Delta_0 \equiv T_{H0} - T_{C0} = 0.2\ K$. Duration of the heat transfer process $t_f = 16000\ s$.

In case of $k = 2 \cdot 10^{-2}\ W/K$ the controlled heat transfer is not oscillatory (Fig. 10a,d,g,j). The optimal control has a jump from $L_{min}$ to $L_{max}$ shortly after the beginning of the process, next has a constant value and finally decreases smoothly toward a new intermediate constant value (Fig. 10j). The current intensity $I$ is always positive and its time variation has a peak (Fig. 10a). Later on the current stabilizes to a very small constant positive value. The temperature $T_H$ reaches a minimum (Fig. 10d), which is slightly below the initial value $T_{C0}$ of the temperature of the body C. The temperature $T_C$ slightly increases above its initial value $T_{C0}$ but later on decreases toward a final constant value (Fig. 10g). The final constant temperatures of the bodies H and C equal each other and are slightly below the initial temperature $T_{C0}$. This requires explanations. The final current intensity does not vanish and the inductance value in the final stages of the process ranges between $L_{min}$ and $L_{max}$. This may be associated with extraction of internal energy in the Peltier element, which is stored as electromagnetic energy in the inductor. Farther comments on these aspects may be found in [8].

For a conductance value $k = 2 \cdot 10^{-3} \, W/K$, the controlled heat transfer process has a tendency to become damped oscillatory (Fig. 10 b,e,h,k). The control has a jump from $L_{min}$ to $L_{max}$ and a reverse jump from $L_{max}$ to $L_{min}$ later on (Fig. 10k). The up and down jump process is repeated two more times but the time variation of $L$ becomes smoother and smoother. Finally, the control has an oscillatory variation of small amplitude around an average value of about 150 H. The current intensity oscillates from positive to negative values, with two maxima and two minima and at larger time becomes constant at a small positive value (Fig. 10b). The time variation of $T_H$ has a minimum (smaller than $T_{C0}$) followed by a maximum and a less obvious couple of minimum and maximum, before a constant value is reached at larger times (Fig. 10e). The temperature $T_C$ decreases from its initial value $T_{C0}$, next it has a peak and finally it becomes constant in time at a value slightly below $T_{C0}$ (Fig. 10h). This may be associated with storage of electromagnetic energy in the inductor, as explained above.

The controlled heat transfer process is obviously damped oscillatory for a conductance value $k = 2 \cdot 10^{-4} \, W/K$ (Fig. 10c, f, i, l). The optimal control consists of an alternation of up and down jumps between $L_{min}$ and $L_{max}$ (Fig. 10l). The intensity of the current and the temperature $T_H$ oscillate in a damped way (Fig. 10c and Fig. 10f, respectively). The oscillations of $T_C$ are damped faster than those of $T_H$ (compare Fig. 10i and Fig. 10f, respectively. The temperatures of H and C bodies at the end of the process are almost equal each other but $T_H(t_f) < T_C(t_f)$. These two temperatures are slightly smaller than $T_{C0}$.

6.CONCLUSIONS

The complex time dependent energy transfer in a system able to cool down a body below the ambient temperature without using an external reservoir of work is analyzed. The system consists of a cold body, a Peltier element and electric circuit containing an inductor of controllable inductance. We call this system *a selfdriven cooler*.

When the cooling process is not controlled, the temperature of the cooled body decreases in time, reaches a minimum and then increases. The heat transfer process should be stopped when the minimum temperature is reached. The temperature of cooled body becomes smaller than the temperature of the cooler. Enhancing the cooling effect does not necessarily need high inductance values. Faster cooling effects are obtained at smaller constant inductance values.

Results obtained for the new selfdriven cooler proposed in this work are shortly listed. The control is the variable inductance.

1.The optimal control is of bang-bang type. This is very convenient for usage in practice. The minimum temperature of the cooled body is lower in case a selfdriven cooler is used instead of a non-selfdriven cooler (the temperature difference is about three degrees centigrade, for the specific cases treated here).

2.Heat transfer processes shorter than a threshold time are characterized by a constant optimal inductance. The optimal control is of bang-band type when that threshold is exceeded. In that case, the control jump is anterior to the moment when the minimum temperature of the cooled body is reached.

3.The minimum temperature reached by a body depends on its mass. There is however a minimum minimorum temperature (for a mass ranging between 0.01 kg and 0.02 kg, for the specific cases treated here). This is an important result since it puts constraints on the mass of the bodies to be effectively cooled in practice.

4.The cooling effects depend on the properties of the Peltier element. The minimum temperature of the cooled body decreases by increasing the thermal conductance of the Peltier element.

5.Thermal damped oscillations may be obtained during both un-controlled and controlled heat transfer processes, under special circumstances. This implies a very small difference between the initial temperatures of the two bodies in contact and (for instance) a specific range of variation for the conductance of the Peltier element.